\documentclass[sigplan,screen]{acmart}

\AtBeginDocument{%
  }

\setcopyright{acmlicensed}
\copyrightyear{2018}
\acmYear{2018}
\acmDOI{XXXXXXX.XXXXXXX}

\acmConference[Conference acronym 'XX]{Make sure to enter the correct
  conference title from your rights confirmation emai}{June 03--05,
  2018}{Woodstock, NY}

\begin{document}

\title{\textbf{SCADE: Scalable Framework for Anomaly Detection in High-Performance System}}

\author{Vaishali Vinay}
\email{vpapneja@microsoft.com}
\affiliation{%
  \institution{Microsoft Security Research}
  \city{Redmond}
  \state{Washington}
  \country{USA}
}  
\author{Anjali Mangal}
\email{anjalimangal@microsoft.com}
\affiliation{%
  \institution{Microsoft Security Research}
  \city{Redmond}
  \state{Washington}
  \country{USA}
}

\begin{abstract}
  In large-scale environments, such as data centers, the detection of anomalous command-line activity is critical for safeguarding against increasingly sophisticated attacks. Traditional anomaly detection systems often struggle in these settings due to lack of labeled data, and the dynamic nature of legitimate administrative activity. In response, we introduce SCADE: a \textbf{S}calable \textbf{C}ommand-Line \textbf{A}nomaly \textbf{D}etection \textbf{E}ngine. SCADE is designed to identify adversarial patterns and detect Living-Off-The-Land (LOL) attacks by leveraging a dual-layer approach. This approach combines a global analysis layer, which scores command rarity across a broad dataset, with a local analysis layer that captures typical command usage patterns at the user and asset levels. This hybrid framework enhances detection accuracy and adaptability without reliance on labeled data. Initial experimental results demonstrate SCADE’s ability to achieve high signal-to-noise ratios (SNR) and precise anomaly differentiation, making it highly suitable for real-time monitoring to support security and compliance in complex operational environments.
\end{abstract}

\maketitle

\section{Introduction}
Command-line interfaces are integral to modern IT operations, offering a flexible and powerful way for administrators to manage systems, run scripts, and automate tasks. However, this utility also makes command-line environments a prime target for adversaries. In recent years, attackers have increasingly employed "Living-Off-The-Land" (LOL) techniques \cite{crowdstrike2018, SOCRadar2024, Cynet2024}, which involve using legitimate command-line tools \cite{LOLBAS_Project} in unauthorized ways to avoid detection. Tools like PowerShell, certutil, and wmiexec are frequently misused to conduct covert operations, blending in seamlessly with routine administrative activities and bypassing traditional security measures. This trend challenges conventional anomaly detection, as LOL attacks tend to be low-signature, producing minimal detectable indicators and difficult to distinguish from legitimate usage \cite{PaloAltoNetworks_FilelessMalware, Microsoft2018, crowdstrike2023LivingOffTheLand, cyware2023LivingOffTheLand}.\\\\
In complex environments such as data centers, the challenges of detecting anomalous command-line behavior are amplified \cite{KNAPP2011533, ACM_HPC_Security}. Data centers host a wide range of sensitive applications, with command-line activity spanning various user roles, assets, and tasks. This diversity, combined with high volumes of legitimate but unusual commands, makes it difficult for traditional detection systems—which often rely on labeled data or fixed rules—to identify genuine threats accurately. The lack of labeled malicious data exacerbates this issue, as models trained on limited or biased samples may produce false positives, leading to alert fatigue for security teams.\\\\
\textbf{SCADE}—the \textbf{Scalable Command-Line Anomaly Detection Engine}—addresses these challenges with a dual-layered, adaptive framework specifically designed to operate effectively in data centers. SCADE’s detection approach combines a global analysis layer that scores the rarity of commands across extensive datasets with a local analysis layer that builds usage baselines at the user and asset level. This multi-layered approach allows SCADE to distinguish between benign anomalies and potentially malicious actions, without relying on labeled data. Our initial results demonstrate that SCADE’s adaptive detection can not only identify LOL-based threats in data centers but also maintain high signal-to-noise ratios (SNR), thereby enhancing real-time monitoring for security and compliance. This adaptive framework significantly reduces the operational overhead for security teams while ensuring robust defenses against emerging threats in command-line activity.

\subsection{Contributions}

In response to the limitations of traditional threat detection methods in large-scale, label-sparse environments, SCADE is engineered as a novel solution tailored to detect anomalous command-line activities in high computation environments such as data centers. SCADE combines adaptability with high precision to address advanced threats, particularly Living-Off-The-Land (LOL) attacks, and offers a robust framework that scales effectively in complex enterprise settings.
The main contributions of this work are as follows:

\begin{itemize}
    \item \textbf{Dual-Layer Detection Architecture}: We propose SCADE, a scalable command-line anomaly detection engine that integrates \textbf{Global Analysis} with \textbf{Local Analysis} for detecting anomalous command-line behavior. This dual-layer approach differentiates legitimate anomalies (Benign Positives) from actual threats (True Positives), effectively addressing LOL attacks, even in environments with sparse labeled data.
    \item \textbf{Statistical Models for Command-Line Rarity Assessment}: SCADE leverages BM25 and Log Entropy models to identify rare and potentially malicious command-line entries across enterprise datasets. This is the first application of these statistical models to command-line anomaly detection, enhancing detection of both common and rare attack vectors.
    \item \textbf{N-Gram Tokenization for Contextual and Granular Anomaly Detection}: A two-level n-gram approach is employed in SCADE, with 1-gram tokenization capturing unique command elements, such as user identifiers or single command-line terms, while 2-gram tokenization identifies rare combinations. This technique refines anomaly detection by capturing isolated as well as contextual anomalies.
    \item \textbf{Dynamic Thresholding for High Signal-to-Noise Ratio (SNR)}: To minimize false positives, SCADE introduces an adaptive threshold selection that tunes scoring based on recent command-line activity patterns, achieving an SNR of >98\%, a critical requirement for data center environments.
    \item \textbf{Scalability and Robustness for Large-Scale Data Centers}: SCADE is designed to efficiently handle billions of command-line events, making it suitable for data centers where previous EDR and ML-based solutions have encountered scalability challenges.
    \item \textbf{Continuous Compliance Monitoring}: Beyond threat detection, SCADE supports compliance monitoring by flagging non-standard practices, helping organizations ensure alignment with security policies and detect unauthorized activity.
    \item \textbf{Low Dependency on Labeled Data}: SCADE’s unsupervised design reduces the need for labeled data and manual analyst intervention, making it highly suitable for environments where labeled malicious data is limited.
    \item \textbf{Future-Proof and Extensible Design}: SCADE’s architecture is adaptable, allowing for the integration of new features, models, and domain knowledge, enabling it to evolve with emerging threats and novel command-line behaviors.
\end{itemize}
These contributions position SCADE as an innovative and adaptable anomaly detection framework, crafted to meet the dynamic and evolving security demands of enterprise-scale environments.

\section{Related Work}
Our research builds upon advancements in anomaly detection across various fields, focusing on command-line analysis, metadata-driven methodologies, and scalable architectures.  We draw insights from a range of academic and industry contributions relevant to these areas. Despite their advancements, existing methods often face limitations in scalability, dynamic adaptability, or effectively utilizing contextual metadata. SCADE addresses these shortcomings with a modular and extensible framework specifically designed to operate efficiently in high-computation environments while integrating rich metadata for robust anomaly detection. 

\subsection{Command-Line Anomaly Detection}

Numerous frameworks have been developed to detect anomalies in command-line behaviors, targeting a variety of use cases and techniques. Stamp et al. \cite{stamp2022livingoffthelandabusedetectionusing} employed simple NLP techniques, hypothesizing that specific tokens or token combinations are more frequently associated with malicious commands. Junius et al. in their research \cite{Junius2023AnoMark} introduced AnoMark, which leverages Markov Chains to model command-line transitions, providing a probabilistic baseline for anomaly detection. However, its effectiveness diminishes in high-dimensional, dynamic environments where transitions are less predictable. Almgren et al. \cite{1310734} leveraged an active learning framework and Ongun et al. proposed LOLAL (Living-Off-The-Land Anomaly Learner) \cite{10.1145/3471621.3471858}, which combines word embeddings and active learning to detect anomalies in Living-Off-The-Land binaries and scripts (LOLBAS). While these methods integrate machine learning classifiers and tokenization strategies to detect adversarial patterns effectively, they face limitations due to their dependence on labeled data and manual feature engineering, making them less adaptable in dynamic, large-scale environments. \\\\
Recent advances in neural and language models have further enriched this domain. Lin et al. explored the use of transformer-based embeddings \cite{lin2024intrusiondetectionscaleassistance}, leveraging reconstruction and classification-based tuning methods to improve anomaly detection in command-line data. Similarly, the work in \cite{CrowdStrike2024BERT, huang2024cmdcalipersemanticawarecommandlineembedding, bdcc7020060} emphasizes the utility of embedding-based approaches in capturing intricate relationships between command arguments and actions. However, computational scalability remains a concern in high-throughput environments .

Ding et al. in their research \cite{9724456} used deep neural networks to identify malicious UNIX user activity for anomaly detection. Complementing these methods, Zubair et al. \cite{liu2023anomalydetectioncommandshell} leveraged transformer architectures to contextualize shell session interactions, achieving state-of-the-art results for anomaly detection in a shell session. However, their high computational costs and reliance on extensive training data make it less adaptable to resource-constrained or highly dynamic environments .

Graph-based methods also represent an emerging frontier like Filar et al. \cite{filar2020problemchilddiscoveringanomalouspatterns} leverage process-child relationship graphs to capture contextual and relational anomalies within command-line data. While these approaches provide a more holistic perspective, they often require significant preprocessing and computational resources .

SCADE builds upon these diverse methodologies, addressing their limitations with a modular, metadata-driven framework tailored for high-computation environments. Unlike prior methods that either rely on extensive labeled data or focus solely on specific anomaly types, SCADE combines global and local anomaly detection layers to achieve scalability, adaptability, and operational relevance.
\subsection{Metadata-Based Detection Techniques}

The integration of contextual metadata is a critical component in anomaly detection, enhancing both interpretability and accuracy. Metadata features such as process lineage, execution times, user IDs, and parent-child process relationships have been extensively explored in prior frameworks \cite{milajerdi2019holmesrealtimeaptdetection, 10.5555/2442626.2442634, Ma2016ProTracerTP}. Additionally, efforts to correlate alert data for reducing noise, such as alert management \cite{milajerdi2019holmesrealtimeaptdetection} and improving the efficacy of alerts \cite{Hassan2019NoDozeCT}, demonstrate the importance of contextual information. However, deploying these approaches in real-world operational settings presents significant challenges. Solutions must account for diverse environmental factors, including industry norms, customer behavior, and organizational dynamics. Without proper adaptation to these factors, detectors may become overly sensitive to specific attack patterns, resulting in frequent false positives. \\\\
Filar et al. introduce the \textit{PROBLEMCHILD} framework \cite{filar2020problemchilddiscoveringanomalouspatterns}, which identifies anomalous patterns through the analysis of parent-child process relationships. By leveraging graph-based structures and pattern-matching techniques, it effectively uncovers malicious process chains. Similarly, Das et al. and Huang et al. in their proposals introduce systems \cite{7299317, huang-etal-2024-cmdcaliper} that emphasizes the role of semantic metadata analysis in real-time malware identification. This approach employs efficient feature extraction techniques to leverage contextual information, enabling rapid and accurate detection in dynamic environments.\\\\
Despite these advancements, most existing methods struggle with adaptability in dynamic, high-computation settings. Their reliance on static thresholds and predefined rules limits their ability to detect subtle or evolving deviations in operational contexts.\\
SCADE addresses these limitations by integrating metadata as a foundational component of its dual-layer analysis. Its scalable architecture unifies metadata features across global and local detection layers, enabling dynamic thresholding and robust anomaly detection. This approach allows SCADE to overcome the constraints of static or rule-based systems, ensuring reliable detection in diverse and evolving environments.
\subsection{Scalability in Anomaly Detection }

Scalability is a critical challenge in high-computation environments, where anomaly detection systems must analyze vast telemetry datasets without sacrificing performance. Research in this area has primarily focused on individual aspects of anomaly detection or dimensionality reduction. Comprehensive surveys \cite{Liu2020ScalableAnomalyDetection,
AGRAWAL2015708,
akoglu2015graph,
chandola2009anomaly,
hodge2004survey} have explored various anomaly detection techniques, offering valuable insights and highlighting unresolved issues. Dimensionality reduction, a key enabler of scalable anomaly detection, has also been extensively reviewed \cite{sorzano2014surveydimensionalityreductiontechniques}. Additional studies \cite{gama2010knowledge,
gupta2013outlier,
heydari2015detection,
inbook,
7724921} have examined challenges in detecting anomalies and managing high-dimensional data. Despite these efforts, limited research addresses the dual challenge of anomaly detection and high-dimensionality in a unified framework, leaving a significant gap in approaches to scalability. SCADE bridges this gap by combining dimensionality reduction with distributed processing, delivering a scalable framework capable of handling large-scale, high-dimensional data while maintaining high detection accuracy.
\section{Problem Formulation}

High-computation environments, such as data centers and cloud platforms, pose unique challenges for detecting anomalous command-line activities due to the sheer scale, diversity, and operational complexity of command executions. Traditional methods for anomaly detection struggle in these settings for the following reasons:

\begin{enumerate}
    \item \textbf{Lack of Labeled Data for Model Training}: High-computation environments typically lack comprehensive, labeled datasets for command-line activities. With minimal known instances of malicious command sequences, traditional supervised learning approaches cannot reliably detect anomalies or adapt to emerging threats.
    \item \textbf{Complexity of Living-Off-the-Land (LOL) Techniques}: Adversaries frequently employ LOL techniques, leveraging legitimate tools and commands to evade detection. Differentiating between benign, non-standard usage and genuinely malicious actions requires advanced behavioral analysis that accounts for both command structure and context.
    \item \textbf{Volume and Variety of Command-Line Activity}: The sheer volume of command-line events and the diversity of commands across users, processes, and devices create a high Signal-to-Noise Ratio (SNR) environment, overwhelming conventional detection systems with frequent false positives. This is especially challenging in dynamic systems where command patterns evolve continuously.
    \item \textbf{Challenges in Parameter and Context Anomaly Detection}: Detecting anomalies in command-line parameters and contextual information—such as user roles, process relationships, and domain affiliations — requires a framework capable of granular and adaptive analysis. Effective detection must account for the variability in how commands are constructed and used across different scenarios.\\
\end{enumerate}
Given these challenges, our goal is to design a scalable, adaptive detection framework that autonomously identifies anomalous command-line behaviors without relying on labeled data. This framework must distinguish between legitimate deviations in command usage (benign positives) and actual threats (true positives) with minimal human intervention, while maintaining high precision and scalability.

\section{Research Methodology}

\begin{figure}
    \centering
    \includegraphics[width=1\linewidth]{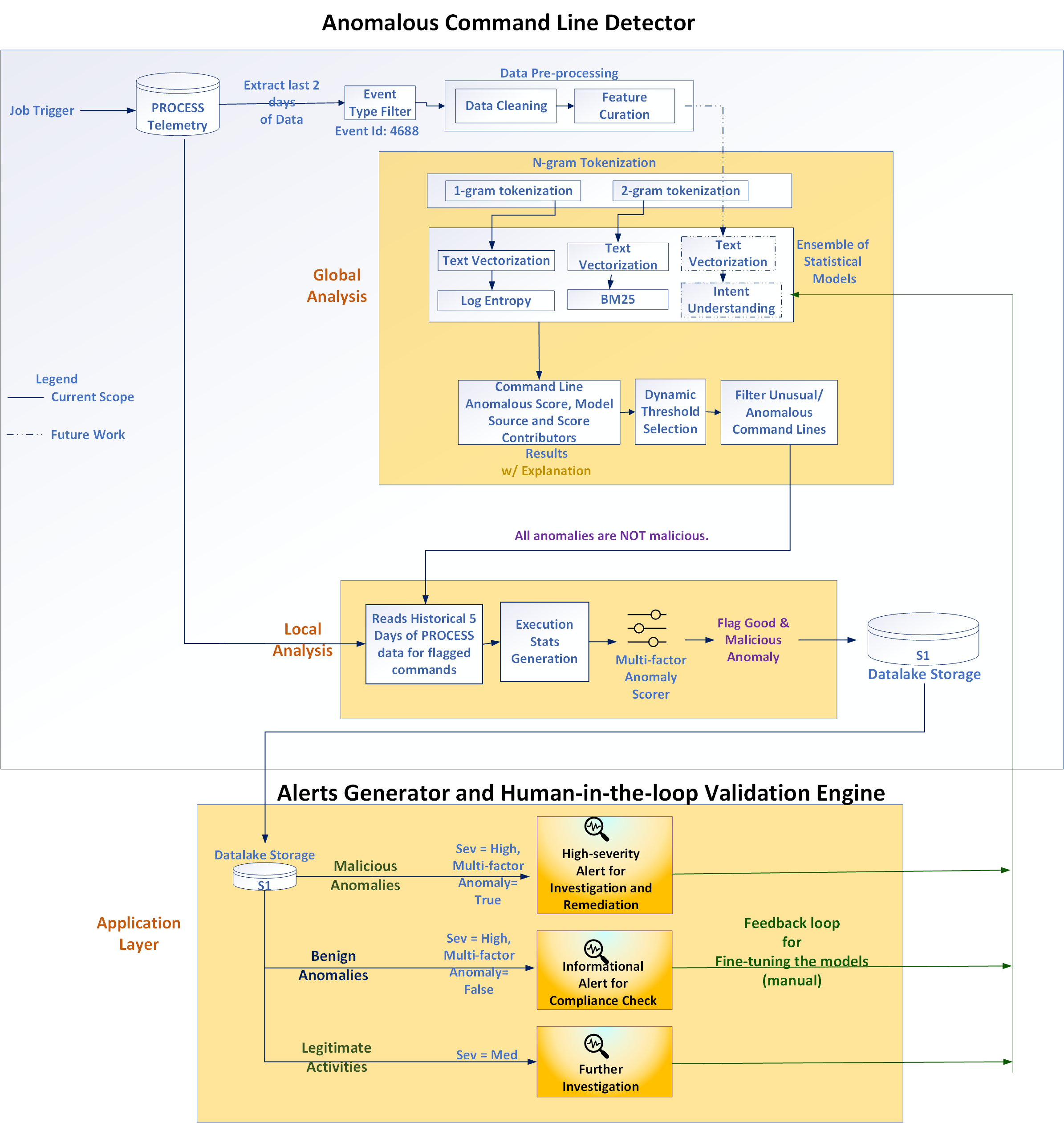}
    \caption{The architecture of the SCADE framework. It highlights the dual-layer analysis approach combining Global and Local Analysis, designed to detect anomalies in command-line activities with precision and contextual awareness.}
    \label{fig:scade_framework}
\end{figure}

The SCADE framework (Figure \ref{fig:scade_framework}) was developed to tackle the challenges inherent in detecting anomalous command-line activities within high-computation environments. In Section 4.1, we establish key intuitions about unusual command-line behavior, detailing what can be considered potentially malicious. Section 4.2 covers the data processing, feature extraction, and curation strategies that transform raw command-line data into structured, analyzable input. Section 4.3 introduces SCADE’s dual-layer architecture, which combines Global and Local Analysis to detect anomalies with precision and context at scale. Finally, Section 4.4 discusses the preliminary evaluation of SCADE.

\subsection{Understanding Malicious Command-Line Behavior }

The detection of malicious command-line behavior hinges on differentiating benign anomalies from actions that may signal potential threats. In high-computation environments, legitimate command-line operations can be complex and varied, often performing critical administrative or computational tasks. However, malicious actors exploit these command-line interfaces by executing "Living-Off-The-Land" (LOL) attacks, where benign utilities are used to conduct harmful activities, enabling evasion of traditional security measures.
Key behaviors that indicate potentially malicious command-line activity include abnormal usage patterns, unexpected command sequences, and rare combinations of command parameters that deviate from typical operational behavior. Table 1 represents the list of key intuitions considered for this work.

\begin{table}
\centering

\begin{tabular}{|l|} \hline 

\textbf{Unusual Command Line Types} \\ \hline 

Small variation in target path or file name \\ \hline 

Unusual combinations of parameters in command lines \\ \hline 

An asset executing a command it is not supposed to run \\ \hline 

Command lines triggered by unexpected parent processes \\ \hline 

Command lines triggering unexpected child processes \\ \hline 

Communication with unusual/malicious IP addresses \\ \hline 

An unusually high number of executions from a given asset\\ \hline 

Malicious intent within command lines \\ \hline

\end{tabular}
\caption{Possible Types of Anomalous Behavior }
\label{tab:my_table}
\end{table}

\subsection{Metadata Considerations in Anomaly Detection  }
In SCADE’s framework, metadata considerations play a critical role in differentiating typical command-line usage from anomalous or potentially malicious behavior. By integrating metadata attributes into the anomaly detection process, we enhance SCADE's ability to recognize deviations that often indicate adversarial tactics, techniques, and procedures (TTPs). The key metadata attributes and their contributions are summarized below:

\begin{enumerate}
    \item \textbf{Command Structure and Parameterization}

Metadata related to command structure and specific parameters allow SCADE to identify rare and suspicious command patterns. By analyzing the sequences and combinations of arguments, SCADE can discern unexpected or inconsistent usage that might otherwise appear benign. For example, deviations in arguments passed to well-known utilities, such as \verb|certutil|, signal potential misuse or unauthorized access attempts.
    \item \textbf{User Context Analysis}

Tracking user accounts associated with command execution reveals usage anomalies, particularly when actions are performed by unexpected or unauthorized accounts. For instance, if a low-privileged user account is observed executing commands commonly reserved for administrative users, this is flagged for deeper analysis. This feature is essential for identifying compromised accounts performing atypical activities.
    \item \textbf{Domain and Network Context}

Associating command execution with domain information supports detection of lateral movement and unauthorized domain access. By flagging discrepancies between observed domains and standard usage patterns, SCADE can detect unauthorized domain access that may indicate an attempt to escalate privileges or traverse the network.
    \item \textbf{Parent-Child Process Relationships}

Understanding the relationships between processes provides insight into how commands are invoked, especially when command lines are initiated by unexpected or unassociated parent processes. This metadata aspect is critical for identifying anomalies like suspicious command invocations from browser processes or other non-standard utilities.
\end{enumerate}
 
\subsection{SCADE Framework}
Our framework is organized into four primary stages, each designed to address a specific layer of command-line anomaly detection. This modular architecture guides the process through data collection, comprehensive global and local analysis, and a final result integration phase. Each stage consists of sequentially defined steps (S1–S10) that build on each other to achieve precise and adaptive anomaly detection across high-computation environments. 
\subsubsection{Data Collection and Feature Curation}\hfill\\
\textbf{S1 - Collect Telemetry:} This initial step involves gathering PROCESS telemetry data from command-line activities across the environment. The collected data includes details on command execution, user identity, domain affiliation, parent and child processes, and associated file paths, among other contextual elements. This data provides a comprehensive foundation for detecting potential threats based on both command structures and execution patterns.\newline\\
\textbf{S2 - Data Pre-processing:} Once collected, the raw telemetry data undergoes a structured preprocessing pipeline to focus on relevant events and standardize the input. Initially, we filter the logs to include only entries with Event ID 4688, which captures new process creation events. We then apply standardization to key fields, such as converting text to lowercase and removing extraneous spaces, to ensure uniform formatting. Subsequently, a new field, payload\_items, is constructed by concatenating standardized attributes. This concatenation creates a unified representation that combines critical features of a command-line event into a single entity, allowing the anomaly detection system to evaluate patterns and unusualness in the combined representation of these attributes, which may reveal anomalies not evident when analyzing them individually.
\subsubsection{Global Analysis}\hfill\\
\textbf{S3 - Tokenization:} In this step, we generate \textbf{both 1-gram and 2-gram tokens} from the curated field, \textit{payload\_items}. Creating both types of tokens serves distinct and complementary purposes. \textbf{1-gram tokens are particularly useful for isolating individual anomalies}, such as unusual users, devices, or command-line parameters, each of which can independently indicate abnormal behavior. On the other hand, \textbf{2-gram tokens are used to capture unusual combinations of any two elements within the field} - such as parameter-parameter pair, process-argument pair, or user-device pair that may not seem anomalous on their own but indicate potential threats when appearing together. This \textbf{dual approach provides a more granular analysis}, allowing SCADE to detect both single-parameter irregularities and suspicious multi-parameter patterns, thereby enhancing the accuracy of anomaly detection.\\\\
\textbf{S4 - Text Vectorization:} Next, we use CountVectorizer to convert each token into a numerical format.\\\\
\textbf{S5 - \textbf{Mathematical Formulation of BM25} }\\
BM25 \cite{10.1561/1500000019, okapi_doc, bm25_wiki} is a popular ranking function in information retrieval, widely used in search engines and recommendation engines. It is based on probabilistic retrieval framework and designed to estimate the relevance of document “D” to a query “Q” by examining how often query terms appear within the document length and term frequency. In this research, we introduce the novel application of the BM25 algorithm for detecting anomalies in command-line activities. To the best of our knowledge, this represents the first instance of leveraging BM25 to evaluate the significance of terms that are both rare and contextually important within command-line datasets. The detailed mathematical formulation of the BM25 score computation, as implemented in SCADE, is outlined below. 
\begin{enumerate}
\item \textbf{Calculate Term Frequency Score for Each Token:} The first step involves calculating the term frequency (TF) score for each token “t” within a payload item “d”. This score, denoted as TF(t,d), quantifies the contribution of a term based on its frequency in "d". In (\ref{tf_score_token}) we represent the mathematical formula to calculate term frequency score of each token.

\begin{equation}\label{tf_score_token}
    TF(t,d) = \frac{f(t,d).(k+1)}{f(t,d) + k.(1-b + b.\frac{|d|}{avg(dl)})}
\end{equation}

Here:\\
f(t,d) represents the raw count of how often a token “t” appears in the payload\_item “d”. The value of k controls how quickly the term frequency saturates as it increases, reducing the dominance of frequently occurring components and ensuring that rarer components are not overshadowed. A high value of k (e.g., 2.0) will mean that higher term frequencies will contribute more significantly to the score, while a low value of k (e.g., 0.5) means that the term frequency will quickly saturate and have diminishing returns in the overall score.
 The \textit{b} parameter controls the degree of normalization for document length, determining how much the length of a document influences its relevance score. A high value of \textit{b} (i.e., close to 1) will mean that longer documents are more heavily penalized, while a low value of \textit{b} (i.e., close to 0) means that document length will have little to no impact on the relevance score.\textit{|d|} is the number of tokens in each payload\_item and \textit{avg(dl)} is the average length of payload\_item sequences across the dataset.
 \item \textbf{Calculate Inverse Document Frequency Scores for Each Token:} The IDF score quantifies the rarity of each term “t” across the entire dataset “D”. A rarer term, appearing in fewer documents, is assigned a higher weight. A common term on the other hand receives a lower weight. IDF for a term “t” is computed as defined in (\ref{idf_score_token}):
\begin{equation}\label{idf_score_token}
    IDF(t) = log( \frac{N - n(t) + 0.5}{n(t) + 0.5}+1)
\end{equation}

Here:\\
N is the total number of payload\_item in the dataset and n(t) is the number of payload items containing the term t.\\

This scoring ensures that frequently occurring terms are downweighted, while rare terms are highlighted, making the metric particularly suitable for anomaly detection. A high IDF value signals that the term contributes significantly to the anomaly likelihood. 

 \item \textbf{Calculate and Explain BM25 Score for Each payload\_item:} Finally, the BM25 score combines TF and IDF values to measure the overall relevance and weight of each payload item "d". This score reflects the combined significance of term frequency and rarity, providing a robust metric for anomaly detection in command sequences. The BM25 score for \textbackslash{}( d \textbackslash{}) is calculated as defined in (\ref{bm25_score_command_seq}): 
\begin{equation}\label{bm25_score_command_seq}
    BM25(d) = \sum_{t} IDF(t).TF(t,d)
\end{equation}

\end{enumerate}
\textbf{S6 - Mathematical Formulation of Log Entropy:} The Log Entropy model \cite{log_entropy_infermatic}, , as implemented in SCADE, is a robust technique for weighting the significance of terms within command-line activity data. It leverages the distribution of terms across the dataset to identify their informational value, effectively handling sparse data while emphasizing tokens that are most informative within each payload item. The detailed mathematical formulation of the Log Entropy score computation, as implemented in SCADE, is outlined below.   

\begin{enumerate}
    \item \textbf{Calculate Term Frequency Score for Each Token:} In this step, we calculate the the term frequency of token “t” in payload\_item “d”, denoted as f(t,d), which represents the raw count of how often a token “t” appears in the payload\_item “d”. This step provides the foundational frequency data necessary for subsequent calculations. 
    \item \textbf{Calculate Global Frequency Score for Each Token}: This step calculates the global frequency “f(t)”, representing the number of payload\_items in the dataset “D” in which token "t" appears. This score adjusts the term's weight by accounting for its ubiquity across the entire dataset, ensuring that common terms are downweighted. 
    \item \textbf{Calculate Log Entropy Weight for Each Token:} We then determine the log entropy weight for token “t” in the command line “d”. This weight captures the balance between term frequency and distribution, calculated as defined in (\ref{log_entropy_token}). 
\begin{equation}\label{log_entropy_token}
\text{LogEntropy}(t,d) = 1 + \frac{f(t,d)}{\sum_{d \in D}f(t,d)} \cdot \log \left( \frac{|D|}{1+f(t)} \right)
\end{equation}
Here:\\
f(t,d) is the frequency of the token “t” in payload\_item “d”, f(t) is the global term frequency of token “t”, representing the number of payload\_items in the complete dataset containing token “t” and |D| is the total number of payload\_items in the complete dataset.\\\\
This calculation highlights the uniqueness of a token within the dataset by amplifying less frequent terms that are disproportionately significant. 

\item \textbf{Calculate and Explain Log Entropy Score for each payload\_item: }Finally, the overall Log Entropy score of a payload item is computed by aggregating the weighted contributions of each token. Each token's score reflects its frequency and uniqueness, and the summation across all tokens in a payload item yields its final score. This comprehensive scoring system identifies terms that are both rare and impactful, enabling SCADE to detect critical anomalies in command-line activities. By analyzing these scores, the framework can pinpoint tokens driving the anomaly and facilitate informed decision-making. \\

\end{enumerate}
\textbf{S7 - Dynamic Threshold Selection:} With BM25 and log entropy anomaly scores for each payload item in our dataset, \textbf{the key question becomes: which score to be flagged as true anomaly?} SCADE addresses this with a component called \textbf{dynamic threshold selection}, designed to adaptively set anomaly thresholds for each execution \textbf{based on recent data patterns and statistical analysis}. Unlike fixed thresholds, which often result in high rate of false positives or false negatives due to variations in system behavior - dynamic threshold selection adjusts to current data characteristics, enhancing the framework's flexibility and accuracy. Here’s a detailed breakdown process.
\begin{enumerate}
    \item \textbf{Data Distribution Analysis: }We first \textbf{calculate the statistical properties} of processed data, particularly the \textbf{mean and standard deviation of anomaly scores} within each run. This approach helps \textbf{establish a baseline for what “normal” behavior looks like}, accommodating fluctuations in activities due to load, seasonality, or specific operations.\\
    \item \textbf{Threshold Calculation:} SCADE applies a threshold approach that leverages standard deviations from the mean anomaly score, aligning with the Central Limit Theorem. This allows us to classify anomaly scores into distinct categories as follows.
\begin{itemize}
    \item \textbf{High-severity anomaly:} A command-line activity is flagged as a high-severity anomaly if it has an anomaly score that lies more than two standard deviations from the mean. Using two standard deviations instead of three provides a balanced approach, allowing the system to detect both significant and moderately unusual behaviors.
    \item \textbf{Medium-severity anomaly:} If a score is between 1.5 and 2 standard deviations from the mean, it is considered a medium-severity anomaly.
    \item \textbf{Low-severity anomaly:} Scores within 1.5 standard deviations are treated as lower-severity anomalies or normal behavior.
\end{itemize}

\end{enumerate}
    3. \textbf{Self-adjusting Mechanism}: In each new execution, SCADE recalculates the mean and standard deviation based on last 2 days of command-line activity logs, enabling it to “reset” its understanding of what constitutes normal behavior. This is crucial in cloud and data center environments where usage patterns evolve, new tools are introduced, or workloads fluctuate. In summary, dynamic threshold selection enables SCADE to operate more sensitively and accurately in a data center environment by constantly tuning its detection criteria to reflect the latest activity patterns. This approach contributes significantly to the robustness and efficiency of anomaly detection across large, variable environments like Azure data centers.\\\\
\textbf{S8 - Filter Unusual Anomalies:} From the previous component, we select only high- and medium- severity anomalies. Currently, the anomalies are identified based on either Log Entropy or BM25 anomaly scores, selecting those that meet the criterion in either algorithm. 
From S6, we have identified the rare payload\_Items; but the domain knowledge tells us that \textbf{all anomalies are not malicious}. For example, a test command setup by an admin to run once a year may be rare but not harmful. To filter out such legitimate cases, we conduct a local analysis phase that examines the typical behavior of each user, asset and command line.

\subsubsection{Local Analysis}\hfill\\
\textbf{S9 - Execution Stats Generator}: This component processed payload items flagged as high- or medium- frequency by global analysis section. To incorporate contextual information, it collects historical data of 5 days to calculate detailed statistics per day on command-line executions specific to each user and asset involved with flagged commands. 
This includes metrics such as flagged command’s execution count on associated asset, total number of commands executed on the that asset, total number of distinct assets executing the flagged command and the overall command count executed by the flagged user every day. By establishing a historical baseline, the Execution Stats Generator allows the system to detect deviations from normal behavior.\\\\
\textbf{S10 - Multi-Factor Anomaly Predictor}: This component leverages the Isolation Forest anomaly detection model on the statistics generated by the previous component. By analyzing multiple contextual factors—such as command execution frequency, unique asset usage, and user-specific command patterns—the Isolation Forest model identifies outliers that deviate from the established baseline. This approach allows the system to score anomalies based on multiple attributes rather than a single factor, enhancing detection precision. The combined multi-factor scoring reduces false positives by distinguishing rare but legitimate patterns from genuinely suspicious behavior.
\subsubsection{Combining the Results of Global and Local Analysis}
In this stage, SCADE combines insights from both global and local analysis to determine whether flagged payload items are genuinely malicious (True Positives, TPs) or benign (Benign Positives, BPs). This differentiation process involves multiple conditions based on historical patterns, and contextual data from both analyses. 
\begin{itemize}
    \item \textbf{True Positives: }If a command exhibits high anomaly scores in either Log Entropy or BM25 models from the global analysis while also deviating significantly from typical usage patterns identified in local analysis, are considered true malicious and sent to our customers for investigation and remediation.
    \item \textbf{Benign Positives:} If a command exhibits high anomaly scores in either Log Entropy or BM25 models from the global analysis but is not seen deviating from typical usage patterns identified in local analysis, are considered legitimate (BPs). The local analysis examines historical and contextual usage patterns, such as the frequency of specific commands or processes for a given asset or user. If the command falls within established norms—meaning it is common for that asset or user—it is flagged as benign, even though it may appear anomalous in a broader context. This helps reduce false positives and ensures that only genuinely suspicious activities are escalated. Based on the application, these BPs can also be sent to customers for a review to keep them informed about routine, authorized activities running in the environment.
    \item \textbf{Legitimate Commands:} The commands exhibiting low and medium anomaly scores in both Log Entropy and BM25 models, are considered legitimate and are discarded by the detection framework.
\end{itemize}

\subsection{Evaluation}
A preliminary Proof of Concept (PoC) was conducted using real command-line activity data from Microsoft’s data center to assess the effectiveness of \textbf{SCADE}. To simulate malicious activities, a \textbf{Red Team exercise} was carried out on test nodes, where the team mimicked potential attackers' tactics, techniques, and procedures (TTPs), replicating methods commonly used in sophisticated cyberattacks. The PoC was conducted on data encompassing all the nodes, ensuring comprehensive coverage. The exercise demonstrated that \textbf{SCADE} successfully identified malicious activities with a Signal-to-Noise Ratio (SNR) of 100\%, while effectively minimizing false positives related to authorized administrative tasks. By incorporating simulated attack scenarios, the PoC confirmed that \textbf{SCADE} can efficiently identify anomalies in real-world enterprise environments while maintaining a low false-positive rate. These promising results provide a solid foundation for further extensive experimentation and future refinements to enhance the system's performance. 

\section{Deployment and Impact}

The SCADE framework was deployed in a high-computation enterprise environment of Azure data centers to detect anomalies in their command-line activities. This environment is characterized by large volumes of command-line data generated across 18 million nodes, making traditional detection systems inefficient due to low signal-to-noise ratios.\\\\
\textbf{Deployment Overview:} 
SCADE was integrated into the enterprise’s security monitoring pipeline, where it processed command-line logs in near real-time. The framework was deployed on a distributed architecture to handle the computational demands of high-volume datasets, ensuring scalability and efficiency. \\\\
\textbf{Initial Impact:} 
\begin{enumerate}
    \item \textbf{Improved Detection Precision}: SCADE demonstrated the ability to identify anomalous command lines with a high precision rate. For instance, it rightfully uncovered the malicious commands ran during red team exercise.
    \item \textbf{Unveiling Hidden Risks}: During its deployment, SCADE flagged two long-running certification dump processes  across various assets. These processes, originally created for testing, had been left active for over a year. Their identification and decommissioning reduced unnecessary resource utilization and eliminated potential security blind spots.
    \item \textbf{Operational Efficiency}: By automating anomaly detection, SCADE significantly reduced the manual effort required to analyze command-line activities. Security analysts were able to prioritize high-severity alerts, leading to faster remediation times and more efficient resource allocation.
\end{enumerate}

\section{Discussion and Future Work}

The SCADE framework provides a scalable and adaptive solution for detecting anomalous command-line activities within high-computation environments. Its dual-layer approach—combining both global and local analysis—has demonstrated significant effectiveness in identifying rare but potentially malicious command sequences and maintaining a high signal-to-noise ratio. While SCADE has demonstrated strong performance in our initial experiments, several areas present opportunities for further research and development. \\\\
\textbf{Intent Understanding:} One notable area for future improvement lies in developing a more refined understanding of command intent. While SCADE currently identifies anomalies based on structural characteristics, incorporating an intent-based analysis could provide deeper insights into the objectives behind specific command executions. For instance, NLP techniques could be employed to semantically interpret commands, allowing for more nuanced differentiation between benign, out-of-the-ordinary actions and truly malicious ones. Moreover, sequence analysis models, such as LSTMs or Transformers, could track temporal patterns across command executions to detect unusual workflows associated with known adversarial TTPs (Tactics, Techniques, and Procedures).\\\\
\textbf{Telemetry Expansion: }Another avenue for future development is expanding SCADE's capabilities to handle diverse telemetry sources beyond command-line data, such as API logs or network data, for a more comprehensive view of system behavior. This could lead to enhanced detection of lateral movement and multi-stage attacks, further strengthening SCADE’s robustness.\\\\
\textbf{Active Learning:} In addition, the integration of active learning techniques could provide a practical solution for dynamically adapting to emerging threats. By involving security analysts in reviewing selected high-confidence anomalies, SCADE could incorporate feedback to continuously refine its detection models, ensuring responsiveness to new attack patterns.

\section{Conclusion}

The SCADE (Scalable Command-Line Anomaly Detection Engine) framework introduces an innovative, scalable approach to identifying anomalous command-line activities within high-computation environments, effectively addressing the challenges of detecting rare, potentially malicious behavior at scale. By combining global statistical analysis with local contextual insights, SCADE leverages dual-layer detection to differentiate legitimate anomalies from true threats. This hybrid approach, which incorporates n-gram tokenization, adaptive thresholding, and statistical models such as BM25 and Log Entropy, enables SCADE to achieve a high signal-to-noise ratio while maintaining detection accuracy across diverse command behaviors.\\\\
Our initial experimental results indicate that SCADE maintains a true positive rate above 98\%, demonstrating its capability to significantly reduce false positives while ensuring precise anomaly detection. This performance makes SCADE particularly valuable in large-scale, data-intensive environments where traditional methods struggle to achieve both precision and scalability. Future work on SCADE may include enhancements for tracking temporal patterns and understanding command intent, further advancing its robustness against evolving adversarial tactics.\\\\
The success of SCADE highlights the potential of integrated detection strategies in cybersecurity, setting a foundation for high-fidelity, scalable threat detection. We hope that SCADE’s architecture and methodology will inform further developments in command-line activity monitoring and advance security standards within high-computation settings.

\section{Acknowledgments}

We thank all our colleagues for their collaborative support throughout this research.

\bibliographystyle{ACM-Reference-Format}
\bibliography{ScadeReferences}

\appendix

\end{document}